\documentclass[%
  aip,
  apl,
  floatfix,
  amsmath,
  amssymb,
  reprint,
  superscriptaddress,
]{revtex4-2}

\usepackage[T1]{fontenc}
\usepackage{lmodern}
\usepackage{cmap}
\usepackage{graphicx}
\usepackage{dcolumn}
\graphicspath{{figures/}}

\begin{document}

\title{In situ estimation of local acoustic pressure amplitude by force balancing with a ferrofluid droplet probe}

\author{Seiya Usui}
\affiliation{%
  Graduate School of Engineering Science, The University of Osaka, 1-3 Machikaneyama, Toyonaka, Osaka 560-8531, Japan
}

\author{Yoshiaki Uchida}
\email{y.uchida.es@osaka-u.ac.jp}
\affiliation{%
  Graduate School of Engineering Science, The University of Osaka, 1-3 Machikaneyama, Toyonaka, Osaka 560-8531, Japan
}

\author{Akira Nagakubo}
\affiliation{%
  Graduate School of Engineering, Tohoku University, 6-6-02 Aramaki Aza Aoba, Aoba-ku, Sendai, Miyagi 980-8579, Japan
}

\author{Norikazu Nishiyama}
\affiliation{%
  Graduate School of Engineering Science, The University of Osaka, 1-3 Machikaneyama, Toyonaka, Osaka 560-8531, Japan
}

\date{\today}

\begin{abstract}
Acoustic tweezers enable non-contact manipulation of microscale objects, but quantitative in situ evaluation of the peak local pressure amplitude remains difficult in confined devices.
Conventional hydrophone-based measurements are often limited at the microscale by probe size and installation constraints.
Here, we present a force-balance method in which a trapped ferrofluid droplet serves as a local probe in a standing-wave acoustic field and an externally applied magnetic-field gradient is tuned so that the magnetic force balances the maximum primary acoustic radiation force on the droplet.
From the magnetic force on the ferrofluid droplet, determined at the balance point, we estimate a peak local pressure amplitude of $2.2\times10^{5}$~Pa for 7.2~MHz operation at 10~V$_{\mathrm{pp}}$.
This approach provides a practical route for quantitative in situ characterization of microscale acoustic fields and for setting operating conditions in compact acoustofluidic devices.
\end{abstract}


\maketitle


The precise manipulation of microscale objects---from individual cells to functional particles---is fundamental to significant advances in fields ranging from cell biology to materials science.
Non-contact techniques for trapping and manipulating microscale objects include magnetic~\cite{Pham2021,TapiaRojo2024}, optical~\cite{Ruttley2025}, acoustic~\cite{Surappa2025,Mao2025,Liu2025,Chen2025}, and composite-type tweezers~\cite{Zhang2024,Hong2024,Zhuang2024}.
Since acoustic tweezers can be applied to a wider range of materials than magnetic tweezers, they provide trapping forces that are considerably stronger than those of optical tweezers and do not interfere with other external stimuli, and they have been applied across a wide range of fields~\cite{Dholakia2020}.
For example, single cells, cell colonies, microbubbles, and liquid crystal droplets can be acoustically manipulated to evaluate their physical properties and behaviors~\cite{Kim2004,JegerMadiot2021,Baresch2020,Naveenkumar2024}.
These techniques rely on the acoustic radiation force, which traps objects by standing waves generated in the acoustic field~\cite{Lim2024}.
Single-beam acoustic tweezers provide localized trapping with a focused ultrasound field and do not require a standing-wave node~\cite{Li2026}.
However, a persistent challenge remains: although the acoustic radiation force can be calculated, there is no practical method to measure it in situ with spatial resolution comparable to the size of the trapped microparticles.
For extremely rare samples or complex compositions, accurately measuring the physical properties required for prediction is difficult.
Quantitative measurement of the acoustic radiation force produced by acoustic tweezers is therefore essential.
Thus, although the radiation force is the quantity most directly felt by the particle, what must be known in the device is the corresponding local pressure amplitude of the standing wave.

To use acoustic tweezers predictably, one must precisely know the local pressure amplitude of the standing wave.
This pressure amplitude is influenced by fine structural details of the system, such as the sensitivity and geometry of the electroacoustic transducers.
Consequently, practical applications require an in situ method for measuring the pressure amplitude in each device.
Currently, hydrophones are commonly used to measure pressure amplitude.
Hydrophones can quantitatively measure absolute pressure and offer high sensitivity over a broad frequency range from 1 kHz to 40 MHz~\cite{Harris2023}.
Hydrophones are well-suited to measuring average pressure in spaces larger than 1 mm.
This is often impossible in enclosed microfluidic channels, and even when feasible, the probe is large enough (typically $>$\,100~$\mu$m) to perturb the delicate standing wave it is intended to measure.
In essence, they fail to provide non-invasive, in situ characterization at the microscale.
Overcoming the limitation would enable the precise manipulation of microscale objects.
For example, when trapping biological cells that are easily compressed or deformed, the force range that does not affect their shape should be estimated in advance~\cite{Zeng2022}.
Moreover, by generating standing waves in a microfluidic channel and adjusting the amplitude or frequency, selective particle separation based on size becomes possible~\cite{Qiu2020,Zhao2023}.
An alternative strategy to overcome this long-standing limitation is a force-probe method that balances the acoustic radiation force acting on a microscale object with an externally applied force that can be calibrated independently of the acoustic field.
Probe-based approaches have been explored to characterize acoustic pressure fields at the microscale.
One such approach exploits the balance between the swimming force of motile cells and the acoustic radiation force to characterize standing-wave fields~\cite{Kim2019}.
This approach can process many cells rapidly, but the swimming force varies among individual cells and is difficult to calibrate quantitatively, which can affect the inferred acoustic field.
Calibrated optical traps provide a complementary route to more precise measurements that eliminate biological variability: a well-defined counter-force measures the acoustic radiation force on a single particle and yields the local pressure amplitude~\cite{Lakamper2014}.
Optical-trap-based methods are, however, limited in force range (typically below $\sim$100~pN) and require direct optical access to the sample.
They are therefore unsuitable when the acoustic radiation force on the probe exceeds this range or when the measurement geometry is enclosed.

Among various forces that can be generated independently of the acoustic radiation force, the magnetic force offers unique advantages due to its high selectivity.
When magnetic particles experience magnetic force in a magnetic field gradient, the surrounding diamagnetic medium is essentially unaffected by the gradient.
Since the magnetic permeability of diamagnetic media is close to unity, the medium's effect on the field is negligible, allowing the magnetic force on the probe to be calculated straightforwardly from the applied field gradient.
Superparamagnetic ferrofluid enables probe droplets whose size and magnetic susceptibility can be tuned, so that the applied magnetic force on each droplet in a magnetic field gradient can be calculated accurately.
In principle, this force-balance approach enables non-invasive, quantitative characterization of acoustic fields.

\begin{figure}[!t]
\includegraphics[width=0.82\columnwidth]{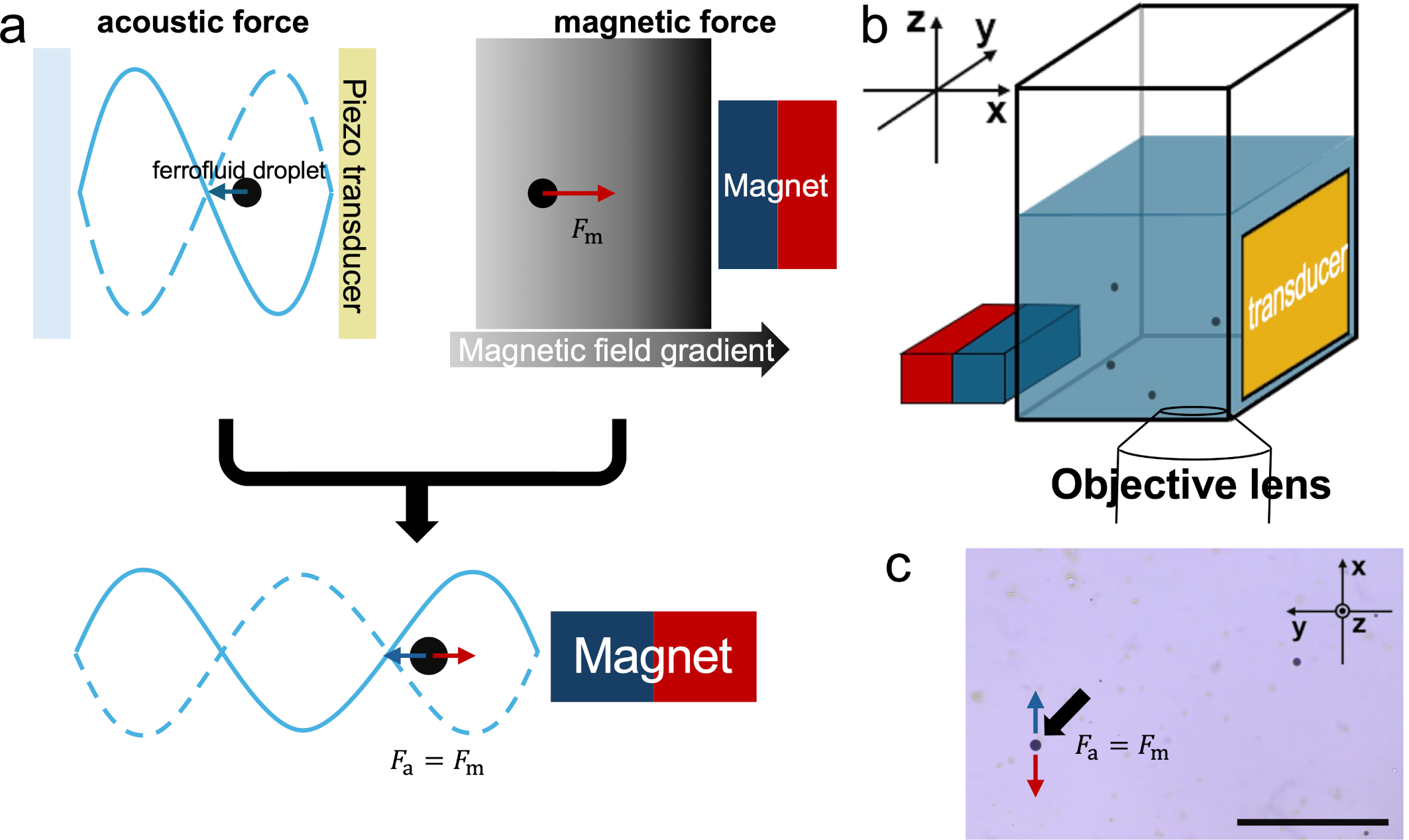}
\caption{Force-balancing approach for in situ peak pressure-amplitude estimation without a perturbative probe. (a) Principle: the acoustic radiation force (blue arrow) and the magnetic force (red arrow) act simultaneously on a ferrofluid droplet trapped in a standing wave; the balanced configuration at threshold is shown together with the permanent magnet. (b) Experimental geometry showing the cuvette, transducer, magnet, and objective lens for imaging. Laboratory axes are labeled in (b) and (c): $x$ along the magnet--cuvette axis (origin at the tip of the magnet), $y$ horizontal and perpendicular to $x$, and $z$ vertical. (c) Representative micrograph at the force-balance threshold ($F_{\mathrm{a}} = F_{\mathrm{m}}$), with blue and red arrows indicating the acoustic and magnetic forces along $x$ (scale bar, 1~mm).}
\label{fig:concept}
\end{figure}

In this work, we implement this concept using a standing-wave plus magnetic-gradient geometry.
A ferrofluid droplet is trapped at a pressure node, and the local pressure amplitude is determined from the force-balance condition, in which the magnetic and acoustic forces balance along the horizontal displacement from that node, as shown in Fig.~\ref{fig:concept}.
This represents, to our knowledge, the first method for non-invasive, in situ estimation of peak local pressure amplitudes without inserting a perturbative probe, such as a hydrophone.

\begin{figure}[!t]
\includegraphics[width=0.90\columnwidth]{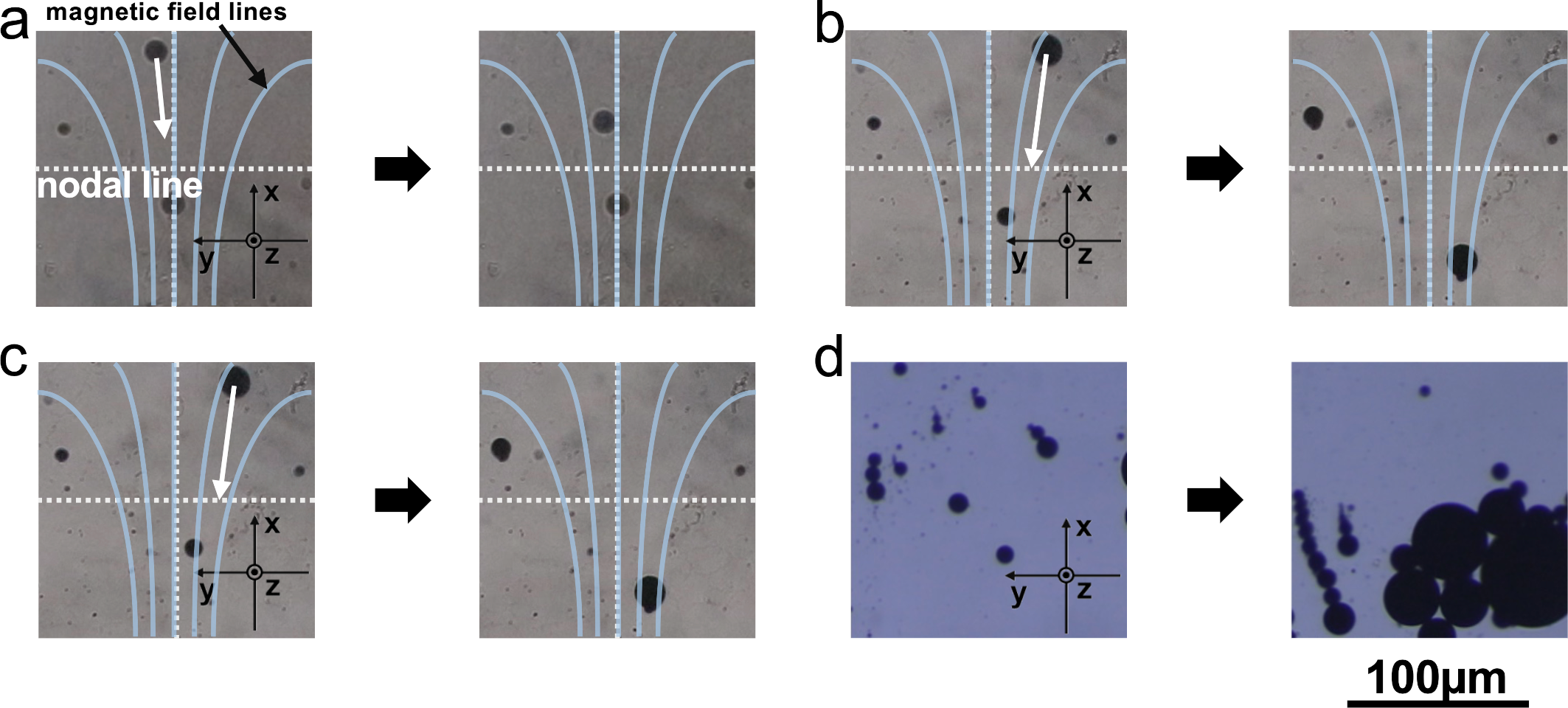}
\caption{Responses of ferrofluid droplets to magnetic and acoustic fields. (a)--(d) The $x$, $y$, and $z$ axes are labeled in each panel (the same laboratory coordinates as in Fig.~\ref{fig:concept}). (a) A ferrofluid droplet abruptly stopped upon approaching another droplet in a magnetic field. (b) A ferrofluid droplet changed course to avoid nearby droplets in a magnetic field. (c) A ferrofluid droplet is attracted to a nearby droplet in a magnetic field. (d) Aggregation of ferrofluid droplets in an acoustic field. White arrows indicate the direction of droplet motion, while blue arrows indicate the direction of the magnetic field. The scale bar is 100~$\mu$m.}
\label{fig:responses}
\end{figure}

To identify experimental conditions suitable for measuring the acoustic radiation force on an individual ferrofluid droplet, we first observed droplet behavior in the as-prepared emulsion at high droplet number density.
The droplets have diameters in the 10--80~$\mu$m range, a size suitable for microscale manipulation.
Initially, we examined the effect of a magnetic field gradient.
When a magnet was placed near the sample, the droplets became magnetized and moved toward it under the influence of the magnetic force.
As droplets approached neighboring droplets, they would abruptly stop, change course, or aggregate, as shown in Fig.~\ref{fig:responses} (see also Supplementary Videos S1--S3).
These behaviors are attributed to dipole--dipole interactions between ferrofluid droplets.
Next, we investigated their behavior in an acoustic field.
When a solution containing many ferrofluid droplets was used, droplet aggregation was observed.
Unlike typical diamagnetic droplets, which align along the pressure nodal line, the ferrofluid droplets did not exhibit such behavior.
This collective behavior is attributed to the secondary acoustic radiation force, which arises when multiple particles are present in an acoustic field.
The secondary acoustic radiation force is an interaction force between particles, mediated by the pressure field and fluid motion induced by acoustic waves.
It becomes pronounced when the inter-particle distance is much smaller than the acoustic wavelength.
Because ferrofluid droplets contain numerous magnetic nanoparticles, they behave as strong acoustic scatterers and are therefore prone to secondary acoustic radiation forces.
These observations indicate that inter-particle interactions significantly affect droplet dynamics under non-dilute conditions.
Therefore, a highly dilute dispersion was employed in this study.

\begin{figure}[!t]
\includegraphics[width=0.90\columnwidth]{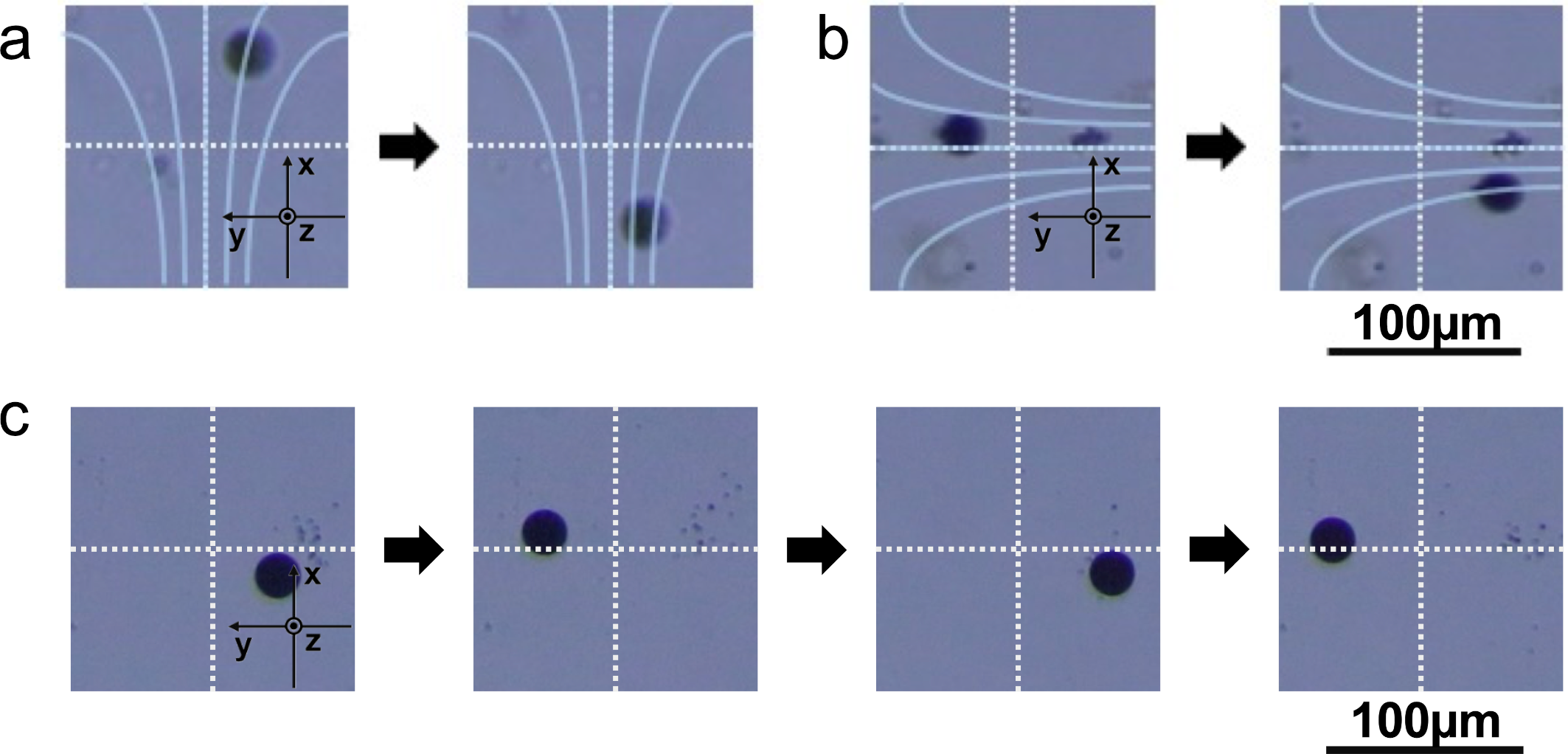}
\caption{Characterization of dual-responsive ferrofluid droplets. The $x$, $y$, and $z$ axes are labeled in each panel (the same laboratory coordinates as in Fig.~\ref{fig:concept}). (a) Pair of optical micrographs (left to right as indicated by the arrow) showing droplet migration while a permanent magnet is brought progressively closer from the bottom of the field of view, with light-blue curves as guides to the eye for the local force landscape; dashed lines mark quadrants (scale bar, 100~$\mu$m). (b) The same arrangement as (a), but with the magnet approached from the right side of the field of view (scale bar, 100~$\mu$m). (c) Four-panel sequence showing reversible repositioning of the droplet when the standing-wave frequency is switched between 7.2~MHz and 7.26~MHz at 10~V$_\mathrm{pp}$ (scale bar, 100~$\mu$m).}
\label{fig:dual}
\end{figure}

Under dilute dispersion conditions, we examined the response of individual ferrofluid droplets to magnetic and acoustic fields to validate their suitability as microscale force probes.
We prepared a dilute emulsion with ferrofluid droplets.
We then systematically characterized their dual responsiveness.
First, to confirm the magnetic response, we brought a magnet close to the droplets, which were immediately attracted along the magnetic field gradient.
Next, we examined their acoustic response in a custom cuvette device equipped with a transducer, as schematically depicted in Fig.~\ref{fig:concept}(b).
Upon generating a 7.2~MHz standing acoustic wave (10~V$_\mathrm{pp}$), the ferrofluid droplet was trapped at a fixed position.
When the frequency was switched to 7.26~MHz while keeping the amplitude constant, the droplet translocated and was trapped at a new position.
This process proved reversible: returning the frequency to 7.2~MHz restored the droplet to the original position.
Because the positions of pressure nodes in a standing wave shift systematically with frequency, this reproducible displacement confirms that the droplet was trapped at a pressure node by the primary acoustic radiation force.
Under these dilute conditions, each droplet can be treated as an isolated force probe, as the inter-droplet spacing is sufficiently large.
Taken together, these results establish that the ferrofluid droplets are reliable, dual-responsive probes for our proposed force-balancing methodology, whose principle is outlined in Fig.~\ref{fig:concept}(a).
Representative optical micrographs documenting these magnetic and acoustic manipulations are presented in Fig.~\ref{fig:dual}.

\begin{figure}[!t]
\includegraphics[width=0.90\columnwidth]{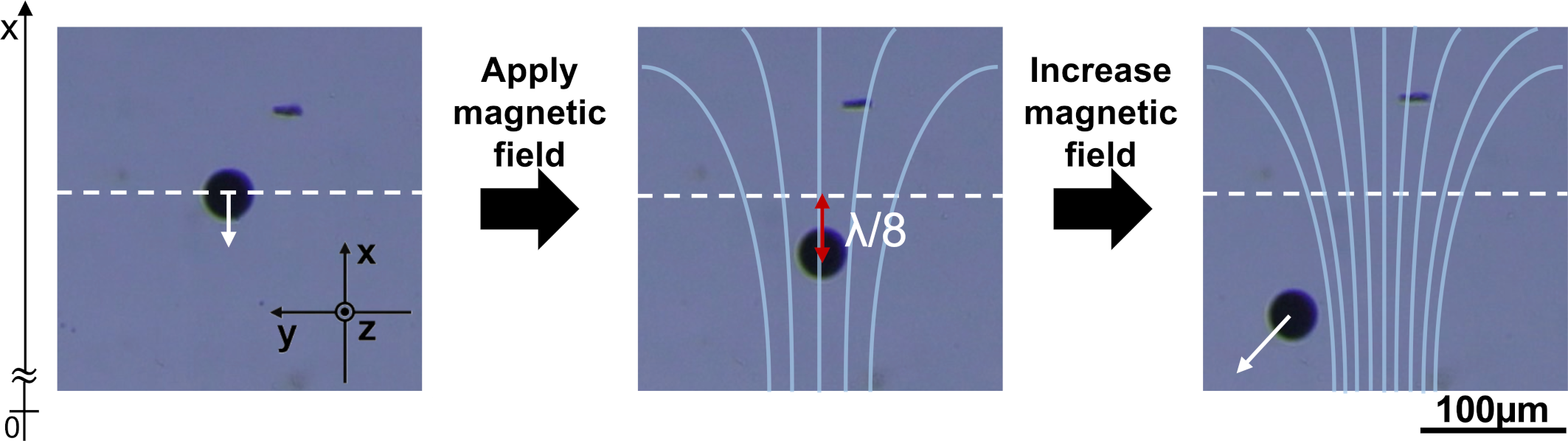}
\caption{Force-balancing behavior of a ferrofluid droplet under an acoustic field and a magnetic field. The $x$, $y$, and $z$ axes are labeled (the same laboratory coordinates as in Fig.~\ref{fig:concept}). The droplet was initially trapped at a pressure node. Applying a magnetic field displaced the droplet from the node, and a balance between the acoustic restoring force and the magnetic force was observed at a displacement of 26~$\mu$m. Upon a further increase in the magnetic field gradient, the droplet moved irreversibly toward the magnet, indicating that the magnetic force exceeded the maximum primary acoustic radiation force. White arrows denote droplet motion. Light-blue curves indicate the magnetic field lines. The dashed line shows the nodal position. Scale bar is 100~$\mu$m.}
\label{fig:forcebalance}
\end{figure}

Our method balances the primary acoustic radiation force with an applied magnetic force.
A ferrofluid droplet was first trapped at a pressure node by a 7.2~MHz standing acoustic wave (10~V$_\mathrm{pp}$).
Then, a permanent magnet was gradually brought closer, and the increasing magnetic force displaced the droplet from the pressure node.
We measured the threshold displacement at droplet escape to be 26~$\mu$m, as shown in Fig.~\ref{fig:forcebalance} (see also Supplementary Video S4).
The corresponding distance $x$ from the tip of the magnet to the droplet at this threshold was 10.0~mm.
Because the target droplet was located on the bottom of the cuvette, only the forces acting in the xy-plane needed to be considered in the force-balance analysis.
Although a permanent magnet generates a three-dimensional field distribution, the calibration records $B$ as a function of distance $x$ from the tip of the magnet to the droplet along the approach axis (Fig.~\ref{fig:concept}), and the gradient $dB/dx$ on this axis enters the force-balance relation (Supplementary Material Sec.~S1).

In applying the force-balance relations below, we adopt three idealizations tied to the experiments described above.
First, the permanent magnet was advanced slowly so that the droplet motion is treated as quasi-static, with viscous drag and fluid inertia regarded as small corrections compared to the magnetic and primary acoustic radiation forces.
Second, the droplet is modeled as a sphere of radius $a$, consistent with the quasi-spherical shapes resolved at the optical magnification used here.
Third, measurements were carried out in the highly dilute regime established earlier, for which secondary acoustic radiation forces between droplets are negligible relative to the primary radiation force, in contrast to the non-dilute behaviors summarized in Fig.~\ref{fig:responses}.
Further discussion of these approximations is provided in Supplementary Material Sec.~S2.
The primary acoustic radiation force in the left-hand side of Eq.~(1) follows the Gor'kov potential for a compressible sphere in an inviscid host fluid~\cite{Gorkov1962}.
Although the emulsion spans droplet diameters of 10--80~$\mu$m, the droplet used in the Fig.~\ref{fig:forcebalance} threshold analysis had a diameter of 30~$\mu$m ($a=15~\mu\mathrm{m}$; Supplementary Material Sec.~S3); at 7.2~MHz the strict long-wavelength condition ($ka \ll 1$) is therefore not necessarily satisfied. Supplementary Material Sec.~S4 shows that evaluating the force with the multipole formulation of Sapozhnikov and Bailey~\cite{Sapozhnikov2013JASA} changes the inferred peak pressure by only approximately 4\% relative to the Gor'kov estimate; the simpler Gor'kov form is therefore retained here for the force-balance calculation, while Gong and Baudoin~\cite{GongBaudoin2021} review equivalent multipole formulations for finite particle size.

Quantitatively, the maximum primary acoustic radiation force is $F = (4\pi a^{3}/3) k E \Phi$, with $E = p^{2}/(4\rho_{m} c_{m}^{2})$ for the peak pressure amplitude $p$~\cite{Dai2021APL,Dai2025SNA}.
At this threshold, the magnetic force balances the maximum primary acoustic radiation force:
\begin{equation}
\frac{4\pi a^{3}}{3}kE\Phi
=-\frac{\chi}{\mu_{0}}\left(\frac{4\pi a^{3}}{3}\right)B\frac{dB}{dx},
\end{equation}
where $a$ is the droplet radius, $k$ is the acoustic wavenumber, $E$ is the acoustic energy density, $\Phi$ is the acoustic contrast factor, $\chi$ is the volume magnetic susceptibility, $\mu_{0}$ is the vacuum permeability, and $B$ is the magnetic flux density.
The acoustic contrast factor $\Phi$ is defined in Supplementary Material Sec.~S4 in terms of the densities and speeds of sound of the droplet and the host medium.
At the threshold distance, the magnetic flux density was 0.0106~T and the magnetic field gradient was $-1.61$~T/m.
Using these values and the force-balance relation, the peak local pressure amplitude was estimated as
\begin{equation}
p=\sqrt{-\frac{4\rho_{m}c_{m}^{2}}{k\Phi}\frac{\chi}{\mu_{0}}\,B\frac{dB}{dx}}
=2.2\times10^{5}~\mathrm{Pa},
\end{equation}
where $\rho_m$ and $c_m$ denote the density and speed of sound of the host aqueous medium, respectively, and $p$ denotes the peak pressure amplitude of the standing wave (not the root-mean-square value).
Propagating the estimated input uncertainties (Supplementary Material Sec.~S5) yields a relative uncertainty of approximately $5\%$ in $p$.
The acoustic energy density $E$ in Eq.~(1) was not evaluated as a separate input; $p$ was obtained directly from the force-balance relation above. Parameter values and their sources for the Fig.~\ref{fig:forcebalance} case are summarized in Supplementary Material Sec.~S3.

In summary, we demonstrated a force-balance approach that estimates the peak local pressure amplitude in a standing-wave acoustic field from the escape threshold of a trapped ferrofluid droplet under a controlled magnetic-field gradient, as shown in Fig.~\ref{fig:forcebalance}.
The escape threshold marks the balance point at which the magnetic force on the droplet equals the maximum primary acoustic radiation force; from the magnetic force determined there together with the measured field parameters, we obtain a peak pressure amplitude estimate of $2.2\times10^{5}$~Pa under the present driving conditions.
Because the droplet acts as an in situ force probe, the evaluation is performed without inserting a hydrophone or otherwise disturbing the acoustic field.
Cross-validation against hydrophone measurements in a less confined geometry remains valuable future work.
Although the present validation was performed in a custom cuvette layout, this capability should help translate device-level driving conditions into physically meaningful acoustic amplitudes for microscale trapping and separation in compact, more strongly confined acoustofluidic systems.

\section*{Supplementary Material}
See the supplementary material for magnetic characterization (Figs.~S1--S3), modeling assumptions, parameter values and uncertainty analysis for the Fig.~\ref{fig:forcebalance} case, the force-balance framework (Gor'kov vs Sapozhnikov), and Videos~S1--S4.

\begin{acknowledgments}
This work was supported in part by the SAKIGAKE Club OU Ecosystem Support Program (in FY2023 and FY2024), and by grants from the Hosokawa Powder Technology Foundation, The Iwatani Naoji Foundation, the Fujimori Science and Technology Foundation, and the Yashima Environment Technology Foundation.
\end{acknowledgments}


%

\end{document}